\documentclass[%
 aip,
 jmp,%
 amsmath,amssymb,
 reprint,%
]{revtex4-1}

\usepackage{graphicx}% Include figure files
\usepackage{dcolumn}% Align table columns on decimal point
\usepackage{bm}% bold math 

\usepackage[cal=boondox]{mathalfa}

\DeclareMathAlphabet\mathbfcal{OMS}{cmsy}{b}{n}

\begin{document}

\preprint{AIP/123-QED}

\title[Nonlinear Cnoidal Waves and Formation of Patterns ...]{Nonlinear Cnoidal Waves and Formation of Patterns and Coherent Structures in Intense Charged Particle Beams}% Force line breaks with \\
%\thanks{Footnote to title of article.}

\author{Stephan I. Tzenov}
\email{tzenov@jinr.ru, \newline {\it Alternative electronic mail}: stephan.tzenov@eli-np.ro}
\affiliation{Baldin Laboratory for High Energy Physics, Joint Institute for Nuclear Research, 6 Joliot-Curie Street, Dubna, Moscow Region, Russian Federation, 141980}

\author{Anton A. Volodin}
\affiliation{Baldin Laboratory for High Energy Physics, Joint Institute for Nuclear Research, 6 Joliot-Curie Street, Dubna, Moscow Region, Russian Federation, 141980}

% \homepage{http://www.Second.institution.edu/~Charlie.Author.}
%\affiliation{%
%Second institution and/or address%\\This line break forced% with \\}%


\begin{abstract} 

The longitudinal dynamics of an intense high energy beam moving in a resonator cavity has been studied in some detail. Through the method of separation of variables and its obvious straightforward generalization, a solution of the Vlasov equation for the distribution function of an intense charged particle beam in the longitudinal direction has been obtained. The thus found Bernstein-Greene-Kruskal (BGK) equilibrium has been utilized to construct stationary wave patterns in the special case when the velocity distribution (energy error distribution) is Maxwellian. These are cnoidal wave patterns, showing rather intriguing and in a sense unexpected analogy between the equilibrium wave patterns in an intense charged particle beam and similar wave clusters originally observed in shallow water.

Based on the hydrodynamic model, fully equivalent to the coupled nonlinear system of the Vlasov equation for the distribution function of an intense beam in the longitudinal direction and the equation for the resonator cavity potential, an amplitude equation in the most general form has been derived. A very interesting and important property of the nonlinear amplitude equation is the fact that it is of hyperbolic type (nonlinear wave equation with complex coefficients) in the entire interval of admissible values for the wave number except
for a single critical point, in which it is of parabolic type (non-linear Schrodinger equation with complex coefficients). 

\end{abstract}

%Use showkeys class option if keyword
\maketitle

\section{\label{sec:intro}Introduction}

Over the years, nonlinear wave phenomena, pattern formation and turbulence have received scant attention in high energy accelerators and storage rings, in part, because of the mathematical difficulty of the subject, but also due to the fact that nonlinear wave motion is commonly associated with a pathological turbulent state of an accelerator that is best to be avoided during operation. In the case of dynamic behavior and propagation of comparatively intense beam, what one mainly interested in is the halo formation around the beam, either in the form of a diffuse cloud, represented mostly by a non equilibrium deviation from Gaussian distribution, or as small droplets of particles that can occur as a kind of phase transition at the edge of the beam due to coherent interaction between oscillation modes or their interaction with trapped particles. Furthermore, it is useful to investigate the formation of an equilibrium state, if such exists, between a wide range of marginally stable wave modes and some weak dissipative mechanisms that can lead to a low-level weak turbulence, which in turn can affect the rate at which the halo population is generated or a weak instability is excited. One might expect that these phenomena are most prevalent in hadron machines due to the weakness of natural radiative damping mechanisms.

Although these topics are mathematically quite complex, there is already a rich literature \cite{davidBOO,davidBOOK,stix,bellan,lieb,piel} and developed advanced methods in the field of plasma physics that deal with these issues, although the interparticle force characteristic of plasma interactions, mostly considered in these references, is due to the space charge alone. At relativistic energies typical of modern accelerators, the interaction between particles is dominated by wall image currents, known in accelerator physics as wakefields, which complicates the nature of the interaction, but can also lead to a greater variety of wave phenomena. If a particle accelerator is operated slightly above its stability limit, due to wakefields the most unstable wave harmonic grows exponentially reaching a saturated, albeit marginally steady state, as the particle distribution in the beam is considerably altered by the wave growth. If the spectrum of unstable wave modes is sufficiently broad, their phases become arbitrarily random, and the interaction between waves and particles leads to diffusion in phase space, known in plasma physics as quasi-linear diffusion \cite{Kennel}. The analogue in charged particle beams, known as the "overshoot" phenomenon has been investigated in Refs. \citenum{Chin} and \citenum{Ng}. Unlike the classical quasi-linear diffusion, the applicability of this model remains unclear owing to the narrow unstable wave spectrum (and hence violation of the condition for random phases), found in the majority of storage rings in operation.

In hadron storage rings, the absence of damping due to synchrotron radiation allows a practically unhindered growth regime, as unstable waves can grow to finite amplitudes which allows a significant fraction of the beam to become trapped in its own wakefield. The resulting wave motion interacts with the trapped particles in such a way that slowly decaying oscillations occur. This phenomenon is known as nonlinear Landau damping in the plasma physics literature \cite{ONeil,oei,onei} (as compared to the linear Landau damping, which is a part of the linear beam response \cite{dawson}). It is to be expected that when a discrete spectrum of unstable waves occurs, as is often the case in hadron machines, this nonlinear Landau damping may play an important role (see e.g. Ref. \citenum{tzenovBOOK}, pages 278--291). 

In all likelihood, the simplest problem to analyze is the longitudinal-only evolution of an intense (coasting or bunched) charged particle beam under the influence of a broadband resonator-type impedance - a practical situation quite common in modern accelerators and storage rings. This model shows a surprisingly wide variety of interesting features, some of which have already been observed experimentally \cite{barton}, and both numerically and theoretically investigated \cite{colestock,Keil}. The case of purely electrostatic impedance is considered in Ref. \citenum{davidson}, where it is shown that the evolution of the linear charge density is described by a Korteweg-de Vries equation. The interaction of a coasting beam with a broad-band resonator-type impedance (of the same type that will be considered in the present article) has been previously studied \cite{tzenovAIP} and a Korteweg-de Vries-Burgers equation to describe the propagation of the linear charge density in the long-wave approximation has been derived.

A wide variety of distinct types of beam equilibria can be detected due to the collective (nonlinear) interaction between the beam particles and the self-consistent resonator waves, induced by the beam itself. Solutions describing analogous types of wave-particle equilibria [Bernstein-Green-Kruskal (BGK) regimes] are well known in plasma physics \cite{bgk} for quite some time. In a nonlinear steady state, coherent structures of arbitrary shape can be formed, which to a significant extent depends on the type (Gaussian, Cauchy, etc.) of initial velocity distribution.

The purpose of the present article is to apply techniques widely used in plasma physics to the study of nonlinear waves and the formation of patterns and coherent structures in intense charged particle beams, which are in close analogy to the well-known BGK plasma modes. In contrast to a previous article \cite{tzenovAIP}, where the propagation of an intense beam was studied in the long-wave approximation (small wave numbers), here this limitation is removed and the problem is treated in all its completeness and generality. 

In the following Section \ref{sec:model}, the statement of the problem is presented, as well as some features and basic properties of the underlying model are highlighted. Then, in Section \ref{sec:vlasov}, a simple solution of the Vlasov equation is found using the method of separation of variables, which is generalized later in Appendix \ref{sec:appendixA}. Based on this solution, it is further shown that an intense beam excites nonlinear quasi-periodic and cnoidal waves in its propagation in a broad-band resonator impedance. With the help of the renormalization group method (RG method) in Section \ref{sec:kdvburg} the most general amplitude equation is derived, which describes the evolution of the envelope of the self-consistent resonator field on long spatiotemporal scales. Finally, Section \ref{sec:conclude} is dedicated to conclusions and outlook.

\section{\label{sec:model}Theoretical Model and Basic Equations}

The single particle longitudinal dynamics of a high energy bunched beam is governed by the set of Hamilton's equations of motion \cite{tzenovBOOK}
\begin{equation}
{\frac {{\rm d} z} {{\rm d} \theta}} = u, \quad \qquad {\frac {{\rm d} u} {{\rm d} \theta}} = - {\frac {{\mathcal K} \Delta E_0} {2 \pi \beta_s^2 E_s}} \sin {\left( {\mathfrak K} z + \Phi_0 \right)} + \lambda V, \label{HamilEq}
\end{equation}
where the azimuth $\theta$ along the accelerator circumference is adopted as an independent variable playing the role of time. The dimensionless canonical variables $z$ and $u$ entering Eqs. (\ref{HamilEq}) can be expressed as follows 
\begin{equation}
z = \theta - \omega_s t, \qquad \qquad u = - {\mathcal{K}} {\frac {\Delta E} {\beta_s^2 E_s}}, \label{CanonVar}
\end{equation}
where $\omega_s$ is the angular revolution frequency of the synchronous particle, $\beta_s$ and $E_s$ are the relative velocity and the energy of the synchronous particle, respectively, $\Delta E$ is the deviation from the energy of the synchronous particle, while ${\mathcal{K}} = \alpha_M - \gamma_s^{-2}$ is the so-called slip-phase factor ($\alpha_M$ - momentum compaction factor). The quantity $\Delta E_0$ on the right-hand-side of the second of Eqs. (\ref{HamilEq}) implies the maximum energy gain per turn, $\mathfrak{K}$ denotes the acceleration harmonic and $\Phi_0$ is the initial phase of the accelerating RF field. The dimensionless variable $V$ represents the wakefield voltage induced in a resonator cavity with shunt impedance $\mathcal{R}$, resonant frequency $\omega_R$ and quality factor $Q$. The resonance frequency $\omega_R$ can be expressed as 
\begin{equation}
\omega_R = {\frac {\nu_{01} c} {R_c}}, \label{ResonFreq}
\end{equation}
where $c$ is the velocity of light in vacuum, $\nu_{01} \approx 2.404826$ is the first zero of the Bessel function $J_0 {\left( w \right)}$, and $R_c$ is the cavity radius. The coupling constant $\lambda$ reads as 
\begin{equation}
\lambda = - {\frac {q^2 {\mathcal{R}} {\mathcal{K}} \omega_R} {2 \pi Q \beta_s^2 E_s}} \varrho_0, \label{CoupCoeff}
\end{equation}
where $q$ is the particle charge and $\varrho_0$ is the uniform beam density distribution in the thermodynamic limit. 

The voltage variation per turn $V {\left( z; \theta \right)}$ satisfies the following equation 
\begin{equation}
\partial_z^2 V - 2 \gamma \partial_z V + \omega^2 V = - \partial_z I, \label{VoltageEq}
\end{equation}
where the beam current $I$ can be expressed as 
\begin{equation}
I {\left( z; \theta \right)} = \int {\rm d} u {\left( 1 + u \right)} f {\left( z, u; \theta \right)}. \label{Current}
\end{equation}
and $f {\left( z, u; \theta \right)}$ is the beam distribution function in the longitudinal direction. Here and in what follows partial derivative with respect to an indicated variable $w$ will be denoted as $\partial_w$. The attenuation increment $\gamma$ and the dimensionless frequency $\omega$ in Eq. (\ref{VoltageEq}) are given by 
\begin{equation}
\gamma = {\frac {\omega} {2 Q}}, \qquad \quad \qquad \omega = {\frac {\omega_R} {\omega_s}}. \label{ResonPar}
\end{equation}
Details regarding the derivation of the above cavity impedance model can be found in Ref. \citenum{Keil} (see also Ref. \citenum{Keil1}). A much simpler and straightforward derivation of the model can be carried out using the theory of equivalent electrical RLC circuits \cite{griff}. For comparison see also Ref. \citenum{heifets}.

With due account of the Hamilton's equations of motion (\ref{HamilEq}), the Vlasov equation for beam distribution function in the longitudinal direction can be written as follows 
\begin{equation}
\partial_{\theta} f + u \partial_z f + {\mathcal{V}} \partial_u f = 0, \label{Vlasov}
\end{equation}
where 
\begin{equation}
{\mathcal{V}} = \lambda V - {\frac {{\mathcal K} \Delta E_0} {2 \pi \beta_s^2 E_s}} \sin {\left( {\mathfrak K} z + \Phi_0 \right)}. \label{GenVolt}
\end{equation}
In addition, the voltage variation per turn $V {\left( z; \theta \right)}$, the beam current $I {\left( z; \theta \right)}$ and the longitudinal distribution function $f {\left( z, u; \theta \right)}$ entering the above equations have been rescaled from their actual values $V_a {\left( z; \theta \right)}$, $I_a {\left( z; \theta \right)}$ and $f_a {\left( z, u; \theta \right)}$, respectively, as follows 
\begin{equation}
V_a = 2 q \omega_s \varrho_0 \gamma {\mathcal{R}} V, \qquad I_a = q \omega_s \varrho_0 I, \qquad f_a = \varrho_0 f. \label{Scaling}
\end{equation}
The system of self-consistent equations (\ref{VoltageEq}), (\ref{Current}) and (\ref{Vlasov}) represents a nonlinear system of Vlasov-Maxwell type, and will constitute the starting point for our systematic analysis in the subsequent exposition.

\section{\label{sec:vlasov}Solution of the Vlasov Equation}

\subsection{\label{sec:vlasovmax}Maxwell-Boltzmann Solution of the Vlasov Equation}

Perhaps, the simplest and most efficient approach to solve the Vlasov equation (\ref{Vlasov}) is the separation of variables method, which can be expressed in the following representation
\begin{equation}
f {\left( z, u; \theta \right)} = g {\left( u \right)} \Psi {\left( z; \theta \right)}, \label{SepVar}
\end{equation}
of the longitudinal distribution function. The continuity equation 
\begin{equation}
\partial_{\theta} \Psi + v_0 \partial_z \Psi = 0, \label{Continuity}
\end{equation}
implies a single equation for the yet unknown function $\Psi {\left( z; \theta \right)}$, where 
\begin{equation}
v_0 = {\left[ \int {\rm d} u g {\left( u \right)}  \right]}^{-1} \int {\rm d} u u g {\left( u \right)} = {\rm const}, \label{SVConst}
\end{equation}
is just a constant number, depending solely on the type of the velocity distribution. Substitution of the ansatz (\ref{SepVar}) into the Vlasov equation (\ref{Vlasov}) with due account of the continuity equation (\ref{Continuity}) yields 
\begin{equation}
{\frac {\partial_z \Psi} {{\mathcal{V}} \Psi}} = {\frac {\partial_u g} {{\left( v_0 - u \right)} g}} \equiv \sigma_u^{-2} = {\rm const}. \label{SVVlasov}
\end{equation}
The above separation of variables identity (\ref{SVVlasov}) leads to the well-known equilibrium Maxwell-Boltzmann distribution 
\begin{eqnarray}
g {\left( u \right)} = {\frac {1} {\sigma_u {\sqrt{2 \pi}}}} \exp {\left[ - {\frac {{\left( u - v_0 \right)}^2} {2 \sigma_u^2}} \right]}, \nonumber 
\\ 
\Psi {\left( z; \theta \right)} = {\mathcal{Z}} \exp {\left[ {\frac {\phi {\left( z; \theta \right)}} {\sigma_u^2}} \right]}, \quad {\mathcal{V}} {\left( z; \theta \right)} = \partial_z \phi {\left( z; \theta \right)}, \label{MaxwBoltz}
\end{eqnarray}
in the entire phase space, where the normalization constant ${\mathcal{Z}}$ is given by the expression 
\begin{equation}
{\mathcal{Z}}^{-1} = {\frac {1} {2 \pi}} \int \limits_0^{2 \pi} {\rm d} z \exp {\left[ {\frac {\phi {\left( z; \theta \right)}} {\sigma_u^2}} \right]}, \label{NormConst}
\end{equation}
provided the velocity distribution $g {\left( u \right)}$ is taken to be normalized. Once the solution to the Vlasov equation is obtained explicitly, the self-consistent potential $\phi {\left( z; \theta \right)}$ can be determined in a straightforward manner.

In order to determine the coordinate-dependent part $\Psi {\left( z; \theta \right)}$ of the distribution function [equivalently, the generalized potential $\phi {\left( z; \theta \right)}$], we substitute the Maxwell-Boltzmann distribution into the right-hand-side of Eq. (\ref{VoltageEq}) governing the evolution of the self-consistent wakefield voltage. As a result, we obtain an equation for the generalized potential $\phi$, which reads as follows  
\begin{widetext}
\begin{equation}
\partial_z^3 \phi - 2 \gamma \partial_z^2 \phi + \omega^2 \partial_z \phi + {\mathcal{A}}_0 {\left[ {\left( \omega^2 - {\mathfrak{K}}^2 \right)} \sin \Phi - 2 \gamma {\mathfrak{K}} \cos \Phi \right]} = - \lambda {\left( 1 + v_0 \right)} {\mathcal{Z}} \partial_z \exp {\left( {\frac {\phi} {\sigma_u^2}} \right)}, \label{GenPotEq}
\end{equation}
\end{widetext}
where 
\begin{equation}
{\mathcal{A}}_0 = {\frac {{\mathcal K} \Delta E_0} {2 \pi \beta_s^2 E_s}}, \qquad \qquad \Phi = {\mathfrak K} z + \Phi_0. \label{AmplPhase}
\end{equation}
Equation (\ref{GenPotEq}) describes the dynamics of a damped nonlinear oscillator with exponential nonlinearity.

Let us note that for dynamical systems far from equilibrium, the distribution function in the subspace of generalized velocities $u$ is non necessarily Gaussian. The case where it is determined with accuracy up to an a priori known initial normalized distribution is considered in Appendix \ref{sec:appendixA}.

\subsection{\label{sec:cnoidal}Nonlinear Cnoidal Waves in a Broad-Band Resonator Impedance}

The coupling parameter $\lambda$ is usually small, so that it can serve as an expansion parameter to solve Eq. (\ref{GenPotEq}) approximately 
\begin{equation}
\phi {\left( z; \theta \right)} = \phi_0 {\left( z; \theta \right)} + \sum \limits_{m=1}^{\infty} \lambda^m \phi_m {\left( z; \theta \right)}. \label{PertExpand}
\end{equation}
Let us consider the solution of Eq. (\ref{GenPotEq}) perturbatively, with subsequent renormalization of the corresponding amplitude and phase, in the simple case of a high-quality resonator ($\gamma \rightarrow 0$). In zeroth order, we have the particular solution 
\begin{equation}
\phi^{(0)} = {\frac {{\cal A}_0} {{\mathfrak{K}}}} \cos \Phi, \label{CWZeroOrdPart}
\end{equation}
and correspondingly 
\begin{equation}
\Psi^{(0)} {\left( z; \theta \right)} = {\mathcal{Z}} \exp {\left( {\frac {{\cal A}_0} {{\mathfrak{K}} \sigma_u^2}} \cos \Phi \right)}. \label{CWZeroOrdPsi}
\end{equation}
Using the Jacobi-Anger expansion \cite{watson} for modified Bessel functions  
\begin{equation}
e^{w \cos \Theta} = I_0 {\left( w \right)} + 2 \sum \limits_{m=1}^{\infty} I_m {\left( w \right)} \cos m \Theta, \label{JacobiAnger}
\end{equation}
we find 
\begin{equation}
{\mathcal{Z}}^{-1} = I_0 {\left({\frac {{\cal A}_0} {{\mathfrak{K}} \sigma_u^2}} \right)}. \label{NormaConst}
\end{equation}
Here $I_n {\left( w \right)}$ is the modified Bessel function of the first kind. The general zero-order solution can be expressed as 
\begin{equation}
\phi_0 = C \cos \zeta + \phi^{(0)}, \qquad \qquad \zeta = \omega z + {\mathfrak{c}}, \label{CWZeroOrdGen}
\end{equation}
where $C$ and ${\mathfrak{c}}$ are constant (to this end) amplitude and phase, respectively. 

With the zero-order solution in hand, we again use the Jacobi-Anger expansion (\ref{JacobiAnger}) to decompose the right-hand side of Eq. (\ref{GenPotEq}), and select the secular terms giving rise to divergent contribution in the perturbation solution. The secular first-order perturbation equation reads as 
\begin{equation}
{\left( \partial_z^2 + \omega^2 \right)} \phi_1 = - 2 {\left( 1 + v_0 \right)} I_1 {\left({\frac {C} {\sigma_u^2}} \right)} \cos \zeta. \label{CWFirOrdSec}
\end{equation}
To prevent the appearance of secular terms in the first-order solution, we need to renormalize the arbitrary amplitude $C$ and phase ${\mathfrak{c}}$ using a method described in the next Section \ref{sec:kdvburg}. Thus, we obtain the following nonlinear amplitude-phase equations 
\begin{equation}
{\left( \partial_z^2 + \omega^2 \right)} C - {\frac {{\mathfrak{M}}^2} {C^3}} = - 2 {\left( 1 + v_0 \right)} I_1 {\left({\frac {C} {\sigma_u^2}} \right)}, \label{CWAmplEquat}
\end{equation}
\begin{equation}
{\mathfrak{M}} = C^2 {\left( \partial_z {\mathfrak{c}} + \omega \right)}, \label{CWPhaseEquat}
\end{equation}
where ${\mathfrak{M}}$ is an integration constant in the sense of an integral of motion. The amplitude equation (\ref{CWAmplEquat}) is also known as the nonlinear generalization of the Ermakov-Pinney equation. 
%% ==============================================
\begin{figure}
\centering
\includegraphics[width=8.5cm]{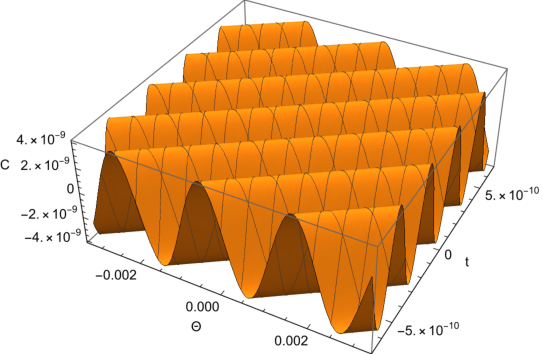}
\caption{\label{fig:quasiper}Evolution in the azimuth $\theta$ and the time $t$ of the nonlinear wave amplitude $C$ according to equation (\ref{CWSolDuffing}). The typical parameters used in the specific calculations are as follows: $\lambda = 5.033 \times 10^{-9}$, $\sigma_u = 3.27 \times 10^{-4}$, ${\cal K} = -0.055$, $\gamma_s = 2.5$.}
\end{figure} 
%% ==============================================

First, we consider the simple case, where the phase variable is chosen according to the relation ${\mathfrak{c}} = - \omega z + {\mathfrak{c}}_0$, implying that the integral of motion ${\mathfrak{M}} = 0$. Respectively, the amplitude equation (\ref{CWAmplEquat}) simplifies considerably. Taking advantage of the modified Bessel function series expansion \cite{watson} 
\begin{equation}
I_1 {\left( w \right)} = {\frac {w} {2}} \sum \limits_{k=0}^{\infty} {\frac {1} {k! \; {\left( k+1 \right)}!}} {\left( {\frac {w} {2}} \right)}^{2k}, \label{CWModBessel}
\end{equation}
the nonlinear amplitude equation (\ref{CWAmplEquat}) is cast in the form of the classical Duffing equation 
\begin{equation}
\partial_z^2 C + p C + q C^3 = 0, \label{CWDuffing}
\end{equation}
with coefficients 
\begin{equation}
p = \omega^2 + {\frac {1 + v_0} {\sigma_u^2}}, \qquad \qquad q = {\frac {1 + v_0} {8 \sigma_u^6}}. \label{CWCoeffic}
\end{equation}
Assuming the obvious initial conditions $C {\left( z_0 \right)} = {\mathcal{C}}$ and $\partial_z C {\left( z_0 \right)} = 0$, the exact solution of the Duffing equation (\ref{CWDuffing}) can be expressed as \cite{tzenovDP}
\begin{equation}
C {\left( z \right)} = {\mathcal{C}} {\rm cn} {\left[ \Omega {\left( z - z_0 \right)}, k \right]}, \label{CWSolDuffing}
\end{equation}
where ${\rm cn} {\left( w, k \right)}$ denotes the elliptic Jacobi cosine function with argument $w$ and elliptic modulus $k$. Moreover, the elliptic harmonic number $\Omega$ and modulus $k$ are given by 
\begin{equation}
\Omega = {\sqrt{p + {\mathcal{C}}^2 q}}, \qquad \qquad k = {\mathcal{C}} {\sqrt{\frac {q} {2 {\left( p + {\mathcal{C}}^2 q \right)}}}}. \label{CWEllipHNM}
\end{equation}
Figure \ref{fig:quasiper} shows the typical behaviour of the nonlinear quasi-periodic wave for realistic beam and resonator cavity parameters.
%% ==============================================
\begin{figure}
\centering
\includegraphics[width=8.5cm]{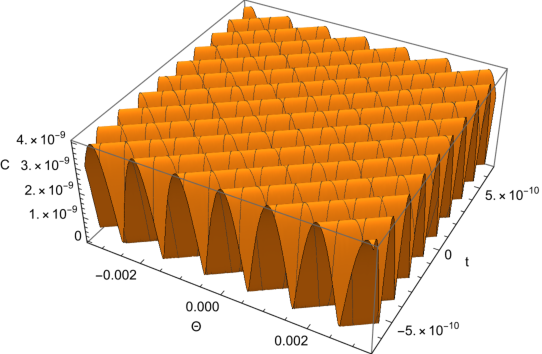}
\caption{\label{fig:cnoidal}Evolution in the azimuth $\theta$ and the time $t$ of the cnoidal wave amplitude $C$ according to equation (\ref{CWSolCnoid}). The typical parameters used in the specific calculations are the same as in Figure \ref{fig:quasiper}.}
\end{figure} 
%% ==============================================

Next, we consider the general case of the Ermakov-Pinney equation for nonzero ${\mathfrak{M}}$. Multiplying both sides of Eq. (\ref{CWAmplEquat}) by $\partial_z C$ and integrating once using the identity 
\begin{equation}
\int {\rm d} w I_1 {\left( w \right)} = I_0 {\left( w \right)}, \label{CWBesselIden}
\end{equation}
we obtain the first integral of the Ermakov-Pinney equation
\begin{equation}
H = {\left( \partial_z C \right)}^2 + {\frac {{\mathfrak{M}}^2} {C^2}} + \omega^2 C^2 + 4 \sigma_u^2 {\left( 1 + v_0 \right)}  I_0 {\left({\frac {C} {\sigma_u^2}} \right)}. \label{CWFirInt}
\end{equation}
Using the series expansion of the modified Bessel function of zero order \cite{watson}
\begin{equation}
I_0 {\left( w \right)} = \sum \limits_{k=0}^{\infty} {\frac {1} {{\left( k! \right)}^2}} {\left( {\frac {w} {2}} \right)}^{2k}, \label{CWModBessel0}
\end{equation}
and confining to the third term in the above expansion, we cast the invariant $H$ in the form 
\begin{equation}
H = {\left( \partial_z C \right)}^2 + {\frac {{\mathfrak{M}}^2} {C^2}} + p C^2 + {\frac {q} {2}} C^4. \label{CWFirIntAF}
\end{equation}
Acting in a similar manner as the case discussed above, we assume again that $\partial_z C {\left( z_0 \right)} = 0$. Then, it can be verified by direct substitution that the Ermakov-Pinney equation corresponding to the integral of motion (\ref{CWFirIntAF}) possesses an exact solution in the form of a cnoidal wave 
\begin{equation}
C {\left( z \right)} = {\sqrt{\frac {X_2 - k_e^2 X_3 {\rm sn}^2 {\left[ \nu {\left( z - z_0 \right)}, k_e \right]}} {1 - k_e^2 {\rm sn}^2 {\left[ \nu {\left( z - z_0 \right)}, k_e \right]}}}}, \label{CWSolCnoid}
\end{equation}
where ${\rm sn} {\left( w, k \right)}$ denotes the elliptic Jacobi sine function with argument $w$ and elliptic modulus $k$. The frequency $\nu$ and the elliptic modulus $k_e$ are given by the expressions
\begin{equation}
\nu = {\sqrt{\frac {q {\left( {\mathcal{C}}^2 - X_3 \right)}} {2}}}, \qquad \qquad k_e = {\sqrt{\frac {{\mathcal{C}}^2 - X_2} {{\mathcal{C}}^2 - X_3}}}. \label{CWFreqMod}
\end{equation}
The constants $X_2$ and $X_3$ entering the equations above can be expressed as follows
\begin{equation}
X_{2,3} = - {\frac {1} {2}} {\left[ {\mathcal{C}}^2 + {\frac {2 p} {q}} \mp \sqrt{{\left( {\mathcal{C}}^2 + {\frac {2 p} {q}} \right)}^2 + {\frac {8 {\mathfrak{M}}^2} {q {\mathcal{C}}^2}}} \right]}
. \label{CWConstX2X3}
\end{equation}
The analogy with the surface gravity waves on shallow water is quite interesting and impressive. Despite the fact that cnoidal waves have been proposed by Korteweg and de Vries in 1895 and later modified by Benjamin, Bona and Mahony in 1972 in a different context \cite{korteweg,bona}, it is intriguing that they can play an important role in wave generation and propagation in intense charged particle beams.

If the Cauchy distribution (\ref{LorentzDF}) is used instead of the Maxwell-Boltzmann distribution, the right-hand-side of Eq. (\ref{GenPotEq}) can be written as follows 
\begin{equation}
{\rm RHS} (17) = - \lambda {\left( 1 + v_0 \right)} {\mathcal{Z}} \partial_z {\left[ {\left( 1 - {\frac {2 \phi} {\sigma_u^2}} \right)}^{-1/2} \right]}. \label{RHSGenPot}
\end{equation}
If necessary, one could rework all the details of the procedure described above and clarify the equilibrium self-consistent wave patterns in the case of a Cauchy distribution function.

\section{\label{sec:kdvburg}The Nonlinear Wave Equation with Complex Coefficients}

In a number of particular cases, the excessively rich information about the evolution of a given system, which the kinetic approach provides, is not so necessary, and then it is convenient to pass to the hydrodynamic description of the underlying motion. Especially if an exact hydrodynamic closure exists, such a representation proves extremely useful and preferable. Let us consider the case where the distribution function $f {\left( z, u; \theta \right)}$ has constant phase-space density (independent of the canonical variables and the time) within a region bounded by simply-connected boundary curves in phase space and zero phase-space density outside. Since the boundary curves are evolving phase-space trajectories, this model is sometimes called the pulsating water-bag model. It is particularly remarkable because it provides an exact hydrodynamic closure \cite{tzenovBOOK,tzendavid}, which is fully equivalent to the Vlasov equation (\ref{Vlasov}) coupled to the equation for the wakefield voltage variation (\ref{VoltageEq}). The gas dynamic equations read as
\begin{equation}
\partial_{\theta} R + \partial_z {\left( R U \right)} = 0, \label{ContinEqua}
\end{equation}
\begin{equation}
\partial_{\theta} U + U \partial_z U + v_T^2 \partial_z {\left( R^2 \right)} = \lambda V - {\mathcal{A}}_0 \sin \Phi, \label{CurrVel}
\end{equation}
\begin{equation}
\partial_z^2 V - 2 \gamma \partial_z V + \omega^2 V = \partial_{\theta} R - \partial_z R, \label{VoltEq}
\end{equation}
where $R$ and $U$ are the beam particle line density and the current velocity, respectively, expressed according to the relations  
\begin{equation}
R = \int {\rm d} u f {\left( z, u; \theta \right)}, \qquad R U = \int {\rm d} u u f {\left( z, u; \theta \right)}. \label{DensCurr}
\end{equation}
In addition, the quantity $v_T$ is the normalized thermal velocity. In what follows, we shall consider the applied external RF electromagnetic field [the last term on the right-hand-side of Eq. (\ref{CurrVel})]  sufficiently small and not affecting the generation and dynamic evolution of nonlinear wave patterns. This is equivalent to allowing for long bunches, or in other words a quasi-coasting beam to the first approximation. 

Let us introduce the scaling \cite{washimi,debnath} 
\begin{equation}
R \longrightarrow 1 + \epsilon R, \qquad U \longrightarrow \epsilon U, \qquad V \longrightarrow \epsilon V, \label{HydScaling}
\end{equation}
of hydrodynamic variables and resonator cavity voltage, where $\epsilon$ is a formal small parameter to be introduced further, which at the end of all order-of-magnitude calculations will be set to unity. The hydrodynamic equations can be rewritten as follows 
\begin{equation}
\partial_{\theta} R + \partial_z U = - \epsilon \partial_z {\left( R U \right)}, \label{ContEqua}
\end{equation}
\begin{equation}
\partial_{\theta} U + 2 v_T^2 \partial_z R - \lambda V = - \epsilon {\left[ U \partial_z U + v_T^2 \partial_z {\left( R^2 \right)} \right]}, \label{CurVel}
\end{equation}
\begin{equation}
\partial_z^2 V - 2 \gamma \partial_z V + \omega^2 V - \partial_{\theta} R + \partial_z R = 0. \label{VoltaEq}
\end{equation}
Next, we differentiate Eq. (\ref{ContEqua}) with respect to $\theta$ and Eq. (\ref{CurVel}) with respect to $z$, respectively, and manipulate the resulting equations in an obvious manner. As a result, we obtain 
\begin{equation}
{\widehat{\boldsymbol{\Box}}} R - \lambda \partial_z V = \epsilon \partial_z {\left[ \partial_{\theta} {\left( R U \right)} - U \partial_z U - v_T^2 \partial_z {\left( R^2 \right)} \right]}, \label{ResultEq}
\end{equation}
where the d'Alembert operator reads as 
\begin{equation}
{\widehat{\boldsymbol{\Box}}} = 2 v_T^2 \partial_z^2 - \partial_{\theta}^2. \label{DAlembert}
\end{equation}
Combining Eqs. (\ref{VoltaEq}) and (\ref{ResultEq}), we arrive at the basic equation 
\begin{equation}
{\left( {\widehat{\boldsymbol{\Box}}} {\widehat{\mathbf{Z}}} - \lambda {\widehat{\mathbf{D}}} \right)} V = \epsilon {\widehat{\mathbf{D}}} {\left[ \partial_{\theta} {\left( R U \right)} - U \partial_z U - v_T^2 \partial_z {\left( R^2 \right)} \right]}, \label{BasicPerEq}
\end{equation}
to be analysed in what follows. Here, the additional operators are defined as follows 
\begin{equation}
{\widehat{\mathbf{Z}}} = \partial_z^2 - 2 \gamma \partial_z + \omega^2, \qquad {\widehat{\mathbf{D}}} = \partial_z {\left( \partial_{\theta} - \partial_z \right)}. \label{Operators}
\end{equation}

Following the standard procedure of the renormalization group (RG) method \cite{tzenovBOOK,nozaki,shiwa,tzenovDP}, the hydrodynamic quantities together with the resonator cavity voltage are represented as a perturbation expansions 
\begin{eqnarray}
R = \sum \limits_{m=0}^{\infty} \epsilon^m R_m, \qquad \quad U = \sum \limits_{m=0}^{\infty} \epsilon^m U_m, \nonumber
\\ 
V = \sum \limits_{m=0}^{\infty} \epsilon^m V_m. \label{PertExpa}
\end{eqnarray}
The next step in the derivation of the fundamental nonlinear amplitude equation, describing the evolution of an intense beam in a resonator cavity with a relatively small $Q$ factor, consists in expanding Eq. (\ref{BasicPerEq}) in the small parameter $\epsilon$, and obtaining its naive perturbation solution order by order. As a rule, the zeroth order consists in establishing the linear dispersion relation, which defines the fundamental oscillation (wave) modes in the linear approximation. Once the linear dispersion relation is established and the fundamental modes are found, the next order includes regular terms comprised of harmonics and possible combinations of the basic modes in case these are more than one. Usually, at second order secular terms proportional to the fundamental modes appear, which are impossible to be removed with means that might be at hand. Thus, one resorts to their resummation using the prescriptions of the renormalization group (RG) method. We follow here an elegant approach, known as the proto RG operator scheme \cite{nozaki,shiwa,tzenovDP}, which has been proposed in the early 2000s to free as much as possible the standard RG theoretical reduction procedure from the necessity of explicit (in the majority of cases, rather cumbersome) calculation of secular terms. Since the procedure described above involves long and cumbersome but quite standard algebraic manipulations, it is set out in Appendix \ref{sec:appendixB}.

The cornerstone of the proto RG operator approach is the formal second-order secular solution (\ref{AmpEquation}). Here we present only a brief description of the renormalization procedure; the interested reader can consult the references cited above, where further details and other interesting applications of the proto RG operator approach can be found. Since, according to the zeroth-order perturbative solution (\ref{ZOSolV}) the complex amplitude ${\mathcal{A}}$ (regarded as a free parameter) is constant with respect to the variables $z$ and $\theta$, the solution of Eq. (\ref{AmpEquation}) will contain secular (proportional to powers of the spatial and temporal variables) terms, which must be resummed appropriately. The standard procedure of renormalization consists in the first place in the introduction of the free parameters $z_R$ and $\theta_R$ and the renormalized amplitude ${\mathcal{A}}_R$ according to the relation ${\mathcal{A}} = {\mathfrak{F}} {\mathcal{A}}_R$, where ${\mathfrak{F}}$ is the so-called renormalization constant. Omitting the first-order solution, which is regular and thus independent of the free parameters just introduced, we can rewrite Eq. (\ref{PertExpa}) as follows 
\begin{eqnarray}
V {\left( z; \theta \right)} = {\mathcal{A}}_R {\left( z_R; \theta_R \right)} e^{i {\left( K z - \Omega \theta \right)}} \nonumber 
\\ 
+ \epsilon^2 {\left[ P {\left( z; \theta \right)} - P {\left( z_R; \theta_R \right)} \right]} e^{i {\left( K z - \Omega \theta \right)}} + \dots + c.c. \label{RenormCond1}
\end{eqnarray}
Let ${\widehat{\mathbfcal{D}}}_R = {\widehat{\mathbfcal{D}}} {\left( \Omega + i \partial_{\theta_R}, K - i \partial_{z_R} \right)}$ denotes the formal dispersion operator on the left-hand-side of equation (\ref{AmpEquation}) with $z$ and $\theta$ being replaced by $z_R$ and $\theta_R$. Provided that $V {\left( z; \theta \right)}$ should not depend on the renormalization parameters, and acting on both sides of Eq. (\ref{RenormCond1}) appropriately, we obtain 
\begin{equation}
0 = {\widehat{\mathbfcal{D}}}_R V = {\widehat{\mathbfcal{D}}}_R  {\left[ {\mathcal{A}}_R - \epsilon^2 P {\left( z_R; \theta_R {\left| {\mathcal{}}_R \right.} \right)} \right]} e^{i \sigma}. \label{RenormCond2}
\end{equation}
Replacing back the original independent variables and making use of equation (\ref{AmpEquation}), we arrive at the sought for nonlinear amplitude equation 
\begin{equation}
{\widehat{\mathbfcal{D}}} {\left( \Omega + i \partial_{\theta}, K - i \partial_z \right)} {\mathcal{A}} = - {\mathcal{G}} {\left| {\mathcal{A}} \right|}^2 {\mathcal{A}} e^{2 \Gamma \theta}. \label{FinAmpEquat}
\end{equation}
This is the most general amplitude equation describing the formation and evolution on relatively long spatio-temporal scales of coherent patterns and structures in intense beams of charged particles. 

\begin{figure}
\centering
\includegraphics[width=8.5cm]{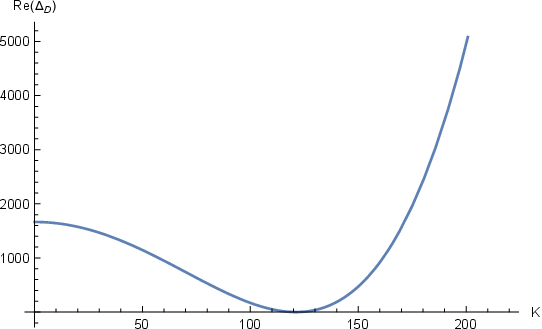}
\caption{\label{fig:discrim}Dependence of the discriminant of Eq. (\ref{NWaveEquat}) on the wave number $K$ in the limit $\gamma \rightarrow 0$. The same parameters as in Figures \ref{fig:quasiper} and \ref{fig:cnoidal} are used in the construction of the graphical dependence $\Delta_D {\left( K \right)}$. It can be seen that the discriminant is positive over the entire range of admissible values of the wave number except for a single point.}
\end{figure}

Strictly speaking, Eq. (\ref{FinAmpEquat}) is a fourth-order partial differential equation with complex coefficients. Usually, the higher order derivatives of third and fourth order, respectively, can be neglected as being significantly smaller than the others. For example, a generic dispersion operator of order $N$ can be expanded in a Taylor series as 
\begin{equation}
{\widehat{\mathbfcal{D}}} {\left( \Omega + i \partial_{\theta}, K - i \partial_z \right)} = \sum \limits_{k=1}^N {\frac {i^k} {k!}} {\left( \partial_{\theta} \partial_{\Omega} - \partial_z \partial_K \right)}^k {\mathcal{D}} {\left( \Omega, K \right)}. \label{TaylorExp}
\end{equation}
Restricting ourselves to second order, we cast the amplitude equation (\ref{FinAmpEquat}) in the form
\begin{eqnarray}
i {\left( {\dot{\mathcal{D}}} \partial_{\theta} - {\mathcal{D}}^{\prime} \partial_z \right)} {\mathcal{A}} - {\frac {1} {2}} {\left( {\ddot{\mathcal{D}}} \partial_{\theta}^2 - 2 {\dot{\mathcal{D}}}^{\prime} \partial_{\theta} \partial_z + {\mathcal{D}}^{\prime \prime} \partial_z^2 \right)} {\mathcal{A}} \nonumber 
\\ 
= - {\mathcal{G}} {\left| {\mathcal{A}} \right|}^2 {\mathcal{A}} e^{2 \Gamma \theta}. \label{NWaveEquat}
\end{eqnarray}
Here, the "raised dot" implies derivative with respect to the wave frequency $\Omega$, while the "prime" stands for derivative with respect to the wave number $K$. The type of the above equation depends on the sign of the so-called discriminant $\Delta_D = {\dot{\mathcal{D}}}^{\prime 2} - {\ddot{\mathcal{D}}} {\mathcal{D}}^{\prime \prime}$. Notice also that in general the second order amplitude equation may be of one type at a specific point in the wave frequency-wave number ${\left( \Omega, K \right)}$-space, and of another type at some other point. All of this depends on the wave number $K$ because the wave frequency $\Omega$ is tied to it by virtue of the dispersion equation (\ref{DispersEq}). Figure \ref{fig:discrim} shows an extremely interesting property of the nonlinear amplitude equation (\ref{NWaveEquat}). The latter is of hyperbolic type in the entire interval of admissible values of the wave number $K$, except for a single point $K_{cr}$, where it is of parabolic type.
%% ==============================================
%% ==============================================

To get a clearer idea of what is actually happening, let us similar to Ref. \citenum{tzenovDP}, introduce the new variable
\begin{equation}
\zeta = z + a \theta, \qquad \quad {\rm where} \qquad \quad a = {\frac {{\dot{\mathcal{D}}}^{\prime}} {{\ddot{\mathcal{D}}}}}, \label{NewVarHyp}
\end{equation}
while the azimuthal variable $\theta$ remains unchanged. Then, the operator in the second brackets of Eq. (\ref{NWaveEquat}) is transformed as 
\begin{equation}
{\ddot{\mathcal{D}}} \partial_{\theta}^2 - {\frac {\Delta_D} {{\ddot{\mathcal{D}}}}} \partial_{\zeta}^2. \label{HypOper}
\end{equation}
Suppose that the discriminant $\Delta_D > 0$ implying that the nonlinear amplitude equation is of hyperbolic type. Indeed, the operator in (\ref{HypOper}) is proportional to the d'Alembert operator, which means that Eq. (\ref{NWaveEquat}) is a {\it nonlinear wave equation with complex coefficients}.

Considering the point $K_{cr}$ at which $\Delta_D = 0$, we define the new variable 
\begin{equation}
\Theta = z + b \theta, \qquad \quad {\rm where} \qquad \quad b = {\frac {{\mathcal{D}}^{\prime \prime}} {{\dot{\mathcal{D}}}^{\prime}}}. \label{NewVarPar}
\end{equation}
while the variable $z$ remains unchanged. Then, the operator in the second brackets of the nonlinear amplitude equation (\ref{NWaveEquat}) simply becomes ${\mathcal{D}}^{\prime \prime} \partial_z^2$. In this case, the nonlinear amplitude equation becomes a {\it nonlinear Schrodinger equation with complex coefficients}.

In summary of all that has been said and worked out above, it can be confirmed that the amplitude equation (\ref{NWaveEquat}) is a wave equation with complex coefficients over the entire range of values for the wave number $K$ except for the critical point $K_{cr}$ where it transforms into a nonlinear Schrodinger equation with complex coefficients. In the limit of $\gamma \rightarrow 0$ the coefficients in the nonlinear Schrodinger equation become real. It is well known that it possesses soliton solutions, which are exact solutions decaying to a background state. In the same nondissipative limit, the coefficients in the nonlinear wave equation are real parameters as well. A similar nonlinear wave equation in a different context was proposed recently in Ref. \citenum{tzenovDP}, where it is shown that its solutions are nonlinear quasi-periodic or cnoidal waves, which were already described in the previous Section.

\section{\label{sec:conclude}Concluding Remarks}

In the present work, we have delineated several different levels of nonlinearity in coherent interactions in intense high-energy beams. Apart from the general academic interest in nonlinear dynamics, for which high-energy beams provide an excellent test-bed, there are a number of other areas and potential implementations where the study of nonlinear wave, wave-particle patterns and coherent structures can find applications in accelerator physics.

We have studied the longitudinal dynamics of an intense high energy beam moving in a resonator cavity. The coupled Vlasov equation for the longitudinal distribution function and the equation for the resonator voltage have been solved by closely following the method of separation of variables, which has been further generalized according to a technique proposed by Karimov and Lewis \cite{Karimov}. The key point of this method consists in the representation of the distribution function as a power series in the resonator potential. Further, self-consistent stationary nonlinear wave patterns have been found in the simplest equilibrium case of Maxwellian distribution in velocity (energy error) space. Very intriguing and somewhat unexpected is the analogy between the equilibrium wave patterns in an intense beam of charged particles and the cnoidal waves originally observed on shallow waters. The main peculiarity of cnoidal waves as compared to standard nonlinear quasi-periodic waves is that the surface elevation of the wave is always positive, which suggests a similarity with the well-known feature of single solitons as a special limit of more general waveforms commonly referred to as cnoidal waves or Turing rolls. One might expect that different travelling waveforms would appear provided the distribution function were of Cauchy type, for example, instead of Maxwell-Boltzmann one.

As a result of the methodical treatment of the hydrodynamic model, fully equivalent to the coupled nonlinear system of Vlasov equation and the equation for the resonator cavity potential, an amplitude equation in the most general form has been derived. It is important to note here that the nonlinear amplitude equation governing the evolution of the hydrodynamic variables on slower spatiotemporal scales is derived without any initial assumptions and/or approximations. A very interesting and important property of the amplitude equation is identified (see Figure \ref{fig:discrim}): it is of hyperbolic type (nonlinear wave equation with complex coefficients) in the entire interval of admissible values of the wave number except for a single critical point $K_{cr}$, in which it is of parabolic type (non-linear Schrodinger equation with complex coefficients). 

A number of pressing issues in the construction and operation of modern hadron colliders come into play in the development of the model described in the present article, and the situation is ripe for careful experimental testing. The benefit of this research is the understanding of the importance of nonlinear waves and coherent structures in the limiting parameters of a given accelerator or storage ring.

\begin{acknowledgments}

Fruitful discussions on topics touched upon in the present article with Prof. Igor Meshkov and Prof. Evgeny Syresin are gratefully acknowledged.

\end{acknowledgments}

\appendix

\section{\label{sec:appendixA}Generalization of the Separation of Variables Method}

A further possible generalization \cite{Karimov} of the separation of variables ansatz (\ref{SepVar}) is suggested by the form of the solution (\ref{MaxwBoltz}) of the Vlasov equation. We assume that the distribution function can be represented as a power-series expansion 
\begin{equation}
f {\left( z, u; \theta \right)} = {\mathcal{Z}} \sum \limits_{n=0}^{\infty} g_n {\left( u \right)} \phi^n {\left( z; \theta \right)}, \label{SepVarGen}
\end{equation}
in the alleged small generalized potential $\phi {\left( z; \theta \right)}$ with yet unknown coefficients $g_n {\left( u \right)}$ depending on the canonical variable $u$ only. The continuity equation 
\begin{equation}
\partial_{\theta} \int {\rm d} u f {\left( z, u; \theta \right)} + \partial_z \int {\rm d} u u f {\left( z, u; \theta \right)} = 0, \label{ContinEq}
\end{equation}
following directly from the Vlasov equation (\ref{Vlasov}) acquires the form 
\begin{equation}
\partial_{\theta} \phi + v_0 {\left( z; \theta \right)} \partial_z \phi = 0. \label{ContinEqu}
\end{equation}
Here 
\begin{equation}
v_0 {\left( z; \theta \right)} = {\dfrac {\sum \limits_{k=0}^{\infty} k A_k \phi^{k-1} {\left( z; \theta \right)}} {\sum \limits_{k=0}^{\infty} k B_k \phi^{k-1} {\left( z; \theta \right)}}}, \label{Funcv0}
\end{equation}
\begin{equation}
A_k = \int {\rm d} u u g_k {\left( u \right)}, \qquad B_k = \int {\rm d} u g_k {\left( u \right)}. \label{CoeffAkBk}
\end{equation}
Next, we substitute the generalized separation of variables ansatz defined according to Eq. (\ref{SepVarGen}) into the Vlasov equation (\ref{Vlasov}). Taking into account (\ref{ContinEqu}), we obtain 
\begin{equation}
{\left( u - v_0 \right)} \sum \limits_{k=1}^{\infty} k g_k {\left( u \right)} \phi^{k-1} {\left( z; \theta \right)} + \sum \limits_{k=0}^{\infty} \phi^k {\left( z; \theta \right)} {\frac {{\rm d} g_k {\left( u \right)}} {{\rm d} u}} = 0. \label{VlasDec}
\end{equation}
In order to determine the unknown functions $g_k {\left( u \right)}$, we raise the assumption that $v_0$ in Eq. (\ref{Funcv0}) does not depend on $\phi {\left( z; \theta \right)}$ and without loss of generality we assume $v_0 = {\rm const}$. This assertion will be proved a posteriori to hold.  

Equating coefficients in front of equal powers of $\phi$ in Eq. (\ref{VlasDec}) yields the following recurrence relation 
\begin{equation}
g_k {\left( u \right)} = {\frac {1} {k}} {\widehat{\boldsymbol{\mathcal{D}}}} g_{k-1} {\left( u \right)}, \label{Recurence}
\end{equation}
where the operator 
\begin{equation}
{\widehat{\boldsymbol{\mathcal{D}}}} = {\frac {1} {v_0 - u}} {\frac{{\rm d}}{{\rm d} u}}, \label{DOper}
\end{equation}
have been introduced. 

The recurrence relation (\ref{Recurence}) yields 
\begin{equation}
g_k {\left( u \right)} = {\frac {1} {k!}} {\widehat{\boldsymbol{\mathcal{D}}}}^k g_0 {\left( u \right)}. \label{RecurSol}
\end{equation}
Thus, we finally arrive at the general solution of the Vlasov equation
\begin{equation}
f {\left( z, u; \theta \right)} = \sum \limits_{k=0}^{\infty} {\frac {\phi^k {\left( z; \theta \right)}} {k!}} {\widehat{\boldsymbol{\mathcal{D}}}}^k g_0 {\left( u \right)} = \exp {\left[ \phi {\left( z; \theta \right)} {\widehat{\boldsymbol{\mathcal{D}}}} \right]} g_0 {\left( u \right)}. \label{VlasovGenSol}
\end{equation}

What remains now is to verify the conjecture $v_0 = {\rm const}$. It suffices to note that \cite{Karimov} 
\begin{widetext}
\begin{equation}
A_k = {\frac {1} {k!}} \int {\rm d} u u {\widehat{\boldsymbol{\mathcal{D}}}}^k g_0 {\left( u \right)} = {\frac {1} {k!}} \int {\rm d} u {\frac {u} {v_0 - u}} {\frac {\rm d} {{\rm d} u}} {\left[ {\widehat{\boldsymbol{\mathcal{D}}}}^{k-1} g_0 {\left( u \right)} \right]} = - {\frac {v_0} {k!}} \int {\frac {{\rm d} u} {{\left( v_0 - u \right)}^2}}  {\widehat{\boldsymbol{\mathcal{D}}}}^{k-1} g_0 {\left( u \right)}, \label{ProofAk}
\end{equation}
\end{widetext} 
\begin{equation}
B_k = - {\frac {1} {k!}} \int {\frac {{\rm d} u} {{\left( v_0 - u \right)}^2}}  {\widehat{\boldsymbol{\mathcal{D}}}}^{k-1} g_0 {\left( u \right)}. \label{ProofBk}
\end{equation}
The proportionality between the two coefficients $A_k$ and $B_k$ 
\begin{equation}
A_k = v_0 B_k, \label{PropAkBk}
\end{equation}
is retained for each $k$ as evident from the last two expressions, implying that the ratio (\ref{Funcv0}) is actually a constant. 

Clearly, the solution (\ref{VlasovGenSol}) is uniquely determined by the generic function $g_0 {\left( u \right)}$. The simplest choice is when $g_0 {\left( u \right)}$ is the Maxwellian distribution (\ref{MaxwBoltz}), that is $g_0 {\left( u \right)}$ itself is an eigenfunction of the operator ${\widehat{\boldsymbol{\mathcal{D}}}}$ with an eigenvalue $\sigma_u^{-2}$ [c.f. equation (\ref{SVVlasov})]. In this case we immediately recover the distribution (\ref{SepVar}) with (\ref{MaxwBoltz}) and the normalization condition (\ref{NormConst}).

The next rather interesting example is when the generic distribution $g_0 {\left( u \right)}$ is the Cauchy probability distribution
\begin{equation}
g_0 {\left( u \right)} = {\frac {\sigma_u} {\pi}} {\frac {1} {{\left( u - v_0 \right)}^2 + \sigma_u^2}}, \qquad \quad \int {\rm d} u g_0 {\left( u \right)} = 1. \label{Lorentzian}
\end{equation}
It can be easily verified (by induction) that the functions $g_n {\left( u \right)}$ can be expressed in terms of $g_0 {\left( u \right)}$ according to the relation 
\begin{equation}
g_n {\left( u \right)} = {\left( {\frac {2 \pi} {\sigma_u}} \right)}^n g_0^{n+1} {\left( u \right)}. \label{LorenInduc}
\end{equation}
Substituting Eq. (\ref{LorenInduc}) back in Eq. (\ref{SepVarGen}), we obtain 
\begin{equation}
f {\left( z, u; \theta \right)} = {\frac {\sigma_u} {\pi}} {\frac {\mathcal{Z}} {{\left( u - v_0 \right)}^2 + \sigma_u^2 - 2 \phi}}. \label{LorentzDF}
\end{equation}
Interestingly enough, the Vlasov distribution function is again a Cauchy probability distribution in momentum space $u$, with a modified by the generalized potential $\phi$, full width at half maximum. The normalizability condition (\ref{NormConst}) can be expressed now as 
\begin{equation}
{\mathcal{Z}}^{-1} = {\frac {\sigma_u} {2 \pi}} \int \limits_0^{2 \pi} {\frac {{\rm d} z} {\sqrt{\sigma_u^2 - 2 \phi}}}. \label{NormCond}
\end{equation}
The two examples, namely the Gaussian and the Cauchy probability distributions discussed above suggest a final generalization. Suppose that the function $g_0 {\left( u \right)}$ is an arbitrary function of $a {\left( u - v_0 \right)}^2 / 2$, that is  
\begin{equation}
g_0 {\left( u \right)} = {\mathcal{F}} {\left( w \right)}, \qquad \qquad w = {\frac{a} {2}} {\left( u - v_0 \right)}^2, \label{ArbFunc}
\end{equation}
where $a = {\rm const}$. Taking into account that 
\begin{equation}
g_n {\left( u \right)} = {\frac {(-1)^n a^n} {n!}} {\left( {\frac {\rm d} {{\rm d} w}} \right)}^n {\mathcal{F}} {\left( w \right)}, \label{GenRecur}
\end{equation}
from Eq. (\ref{SepVarGen}) we readily obtain 
\begin{eqnarray}
f {\left( z, u; \theta \right)} && = {\mathcal{Z}} \sum \limits_{n=0}^{\infty} {\frac {(-1)^n a^n} {n!}} \phi^n {\left( z; \theta \right)} {\left( {\frac {\rm d} {{\rm d} w}} \right)}^n {\mathcal{F}} {\left( w \right)} \nonumber 
\\ 
&& = {\mathcal{Z}} {\mathcal{F}} {\left[ w - a \phi {\left( z; \theta \right)} \right]}. \label{MostGenSol}
\end{eqnarray}
Given the condition (\ref{ContinEqu}), this result is not unexpected and essentially represents the general solution of the Vlasov equation.

\section{\label{sec:appendixB}Derivation of the Generalized Nonlinear Amplitude Equation}

In this Appendix, the details of the derivation of the amplitude equation describing the slow evolution of patterns in an intense beam traversing a resonator cavity with small $Q$ factor, are presented. 

\vspace{5pt} 
{\it Zeroth Order}. 

The zeroth-order basic perturbation equation (\ref{BasicPerEq}) can be written as 
\begin{equation}
{\left( {\widehat{\boldsymbol{\Box}}} {\widehat{\mathbf{Z}}} - \lambda {\widehat{\mathbf{D}}} \right)} V_0 = 0. \label{BasicPEq0}
\end{equation}
The solution of Eq. (\ref{BasicPEq0}) is usually sought in the form of a travelling wave $\sim e^{i \sigma}$ with phase 
\begin{equation}
\sigma = K z - \Omega \theta. \label{Phase}
\end{equation}
The wave number $K$ and the wave frequency $\Omega$ satisfy the dispersion equation 
\begin{equation}
{\mathbfcal{D}} {\left( \Omega, K \right)} = \Box {\left( \Omega, K \right)} Z {\left( K \right)} - \lambda D {\left( \Omega, K \right)} = 0, \label{DispersEq}
\end{equation}
where 
\begin{equation}
\Box {\left( \Omega, K \right)} = \Omega^2 - 2 v_T^2 K^2, \qquad D {\left( \Omega, K \right)} = K {\left( \Omega + K \right)}, \label{DispFunc1}
\end{equation}
\begin{equation}
Z {\left( K \right)} = \omega^2 - K^2 - 2 i \gamma K, \label{DispFunc2}
\end{equation}
The dispersion equation (\ref{DispersEq}) is a quadratic equation in terms of the frequency $\Omega$ as a function of the wave number $K$. Observe that, if $\Omega {\left( K \right)}$ is a solution of Eq. (\ref{DispersEq}), solution is $- \Omega^{\ast} {\left( - K \right)}$ as well. Taking into account this fact, Eq. (\ref{BasicPEq0}) can be readily solved to yield 
\begin{equation}
V_0 = {\mathcal{A}} e^{i {\left( K z - \Omega \theta \right)}} + {\mathcal{A}}^{\ast} e^{-i {\left( K z - \Omega^{\ast} \theta \right)}}. \label{ZOSolV}
\end{equation}
Here
\begin{equation}
\Omega = {\frac {1} {2 Z}} {\left[ \lambda K + {\sqrt{\lambda^2 K^2 + 4 K^2 Z {\left( \lambda + 2 v_T^2 Z \right)}}} \right]}, \label{BasFrequency}
\end{equation}
while the quantity ${\mathcal{A}}$ is an arbitrary complex constant. 

From Eqs. (\ref{ContEqua}) and (\ref{VoltaEq}) for the zeroth-order particle density $R_0$ and current velocity $U_0$, we obtain 
\begin{equation}
R_0 = i {\left( r_0 {\mathcal{A}} e^{i \sigma} - r_0^{\ast} {\mathcal{A}}^{\ast} e^{-i \sigma^{\ast}}  \right)}, \qquad r_0 = {\frac {Z {\left( K \right)}} {\Omega + K}}, \label{ZOSolR}
\end{equation}
\begin{equation}
U_0 = i {\left( u_0 {\mathcal{A}} e^{i \sigma} - u_0^{\ast} {\mathcal{A}}^{\ast} e^{-i \sigma^{\ast}}  \right)}, \qquad u_0 = {\frac {\Omega Z {\left( K \right)}} {K {\left( \Omega + K \right)}}}. \label{ZOSolU}
\end{equation}

\vspace{5pt} 
{\it First Order}. 

The first-order basic perturbation equation (\ref{BasicPerEq}) can be written as 
\begin{equation}
{\left( {\widehat{\boldsymbol{\Box}}} {\widehat{\mathbf{Z}}} - \lambda {\widehat{\mathbf{D}}} \right)} V_1 = {\widehat{\mathbf{D}}} {\left[ \partial_{\theta} {\left( R_0 U_0 \right)} - U_0 \partial_z U_0 - v_T^2 \partial_z {\left( R_0^2 \right)} \right]}, \label{BasicPEq1}
\end{equation}
In order to better navigate the situation, let us write out the right-hand-side of the above equation in a detailed explicit form using the already known zeroth-order solutions for the density and the current velocity 
\begin{equation}
{\left( {\widehat{\boldsymbol{\Box}}} {\widehat{\mathbf{Z}}} - \lambda {\widehat{\mathbf{D}}} \right)} V_1 = {\frac {4i Z^2} {\Omega + K}} {\left( 3 \Omega^2 + 2 K^2 v_T^2 \right)} {\mathcal{A}}^2 e^{2i \sigma} + c.c., \label{BasicPEq11}
\end{equation}
where "$c.c.$" stands for the complex conjugate counterpart. The regular first-order solutions are 
\begin{equation}
V_1 = - {\frac {i Z^2} {\Omega + K}} {\frac {3 \Omega^2 + 2 K^2 v_T^2} {\Box {\left( 3 K^2 + 2i \gamma K \right)}}} {\mathcal{A}}^2 e^{2i \sigma} + c.c., \label{FOrdSolV}
\end{equation}
\begin{equation}
R_1 = r_1 {\mathcal{A}}^2 e^{2i \sigma} + c.c., \label{FOrdSolR}
\end{equation}
\begin{equation}
U_1 = u_1 {\mathcal{A}}^2 e^{2i \sigma} + c.c., \label{FOrdSolU}
\end{equation}
where 
\begin{equation}
r_1 = {\frac {Z^2 Z_2} {2 {\left( \Omega + K \right)}^2}} {\frac {3 \Omega^2 + 2 K^2 v_T^2} {\Box {\left( 3 K^2 + 2i \gamma K \right)}}}, \qquad Z_2 = Z {\left( 2 K \right)}, \label{FOrdSolRr}
\end{equation}
\begin{equation}
u_1 = {\frac {\Omega Z^2} {K {\left( \Omega + K \right)}^2}} {\left[ 1 + {\frac {Z_2} {2}} {\frac {3 \Omega^2 + 2 K^2 v_T^2} {\Box {\left( 3 K^2 + 2i \gamma K \right)}}} \right]}. \label{FOrdSolUu}
\end{equation}

\vspace{5pt} 
{\it Second Order. Derivation of the Amplitude Equation}. 

The second-order basic perturbation equation (\ref{BasicPerEq}) for the hydrodynamic quantities and the resonator cavity voltage read as 
\begin{eqnarray}
{\left( {\widehat{\boldsymbol{\Box}}} {\widehat{\mathbf{Z}}} - \lambda {\widehat{\mathbf{D}}} \right)} V_2 = {\widehat{\mathbf{D}}} {\left[ \partial_{\theta} {\left( R_0 U_1 + U_0 R_1 \right)} \right.} \nonumber 
\\ 
{\left. - \partial_z {\left( U_0 U_1 \right)} - 2 v_T^2 \partial_z {\left( R_0 R_1 \right)} \right]}, \label{BasicPEq2}
\end{eqnarray}
In first-order we encountered only regular terms, which follow the second harmonic pattern (proportional to $e^{2i \sigma}$) of the basic mode in linear approximation. However, the situation in second order is different. Here, two types of terms, resonant (proportional to $e^{i \sigma}$) and regular on the right-hand-side of the second-order equation (\ref{BasicPEq2}) are present. In order to avoid divergent contributions in the perturbation solution, the secular terms must be renormalized away by the elegant renormalization group procedure. It is necessary to represent the right-hand-side of Eq. (\ref{BasicPEq2}) in explicit form using the already known solutions from previous orders and then to isolate the secular contributions. Omitting straightforwardly reproducible calculation’s details, we write down the resonant part of Eq. (\ref{BasicPEq2}) as 
\begin{equation}
{\left( {\widehat{\boldsymbol{\Box}}} {\widehat{\mathbf{Z}}} - \lambda {\widehat{\mathbf{D}}} \right)} V_2 = - {\mathcal{G}} {\left| {\mathcal{A}} \right|}^2 {\mathcal{A}} e^{i \sigma} e^{2 \Gamma \theta}, \label{SeculBasicPEq}
\end{equation}
where 
\begin{eqnarray}
{\mathcal{G}} = K {\left( K + \Omega + 2i \Gamma \right)} {\left[ {\left( \Omega + 2i \Gamma \right)} {\left( r_0^{\ast} u_1 + u_0^{\ast} r_1 \right)} \right.} \nonumber 
\\ 
{\left. + K {\left( u_0^{\ast} u_1 + 2 v_T^2 r_0^{\ast} r_1 \right)} \right]}, \label{GConst}
\end{eqnarray}
and $\Gamma = {\rm Im} {\left( \Omega \right)}$ is the imaginary part of the wave frequency, which is considered to be small. Here we note that the operator on the left-hand-side of Eq. (\ref{SeculBasicPEq}) is actually the dispersion function (\ref{DispersEq}) in which $K$ is replaced by $-i \partial_z$ and $\Omega$ replaced by $i \partial_{\theta}$, respectively. Equation (\ref{SeculBasicPEq}) possesses an exact solution of the form 
\begin{equation}
V_2 {\left( z; \theta \right)} = P {\left( z; \theta \right)} e^{i {\left( K z - \Omega \theta \right)}}. \label{ExactSol}
\end{equation}
Considering that the differential operators acting on the right-hand-side of the above equation are transformed as $\partial_z \longrightarrow \partial_z + i K$ and $\partial_{\theta} \longrightarrow \partial_{\theta} - i \Omega$, the amplitude $P {\left( z; \theta \right)}$ must satisfy the equation
\begin{equation}
{\widehat{\mathbfcal{D}}} {\left( \Omega + i \partial_{\theta}, K - i \partial_z \right)} P = - {\mathcal{G}} {\left| {\mathcal{A}} \right|}^2 {\mathcal{A}} e^{2 \Gamma \theta}. \label{AmpEquation}
\end{equation}
The above equation represents the formal second-order secular solution, which is a subject of renormalization using the renormalization group procedure.

\vspace{10pt}
\hrule

\bibliographystyle{ieeetr}
\bibliography{nonlinwaves.bib}

\begin{thebibliography}{10}

\bibitem{davidBOO}
R.~C. Davidson, {\em Physics of Nonneutral Plasmas}.
\newblock World Scientific, (2001).

\bibitem{davidBOOK}
R.~C. Davidson, {\em Methods in Nonlinear Plasma Theory}.
\newblock Academic Press, New York, (1972).

\bibitem{stix}
T.~H. Stix, {\em The Theory of Plasma Waves}.
\newblock American Institute of Physics, New York, (1992).

\bibitem{bellan}
P.~M. Bellan, {\em Fundamentals of Plasma Physics}.
\newblock Cambridge University Press, (2012).

\bibitem{lieb}
{\rm Michael A. Lieberman, Allan J. Lichtenberg}, {\em Principles of Plasma Discharges and Materials Processing}.
\newblock John Wiley and Sons, (2005).

\bibitem{piel}
{\rm Alexander Piel}, {\em Plasma Physics: An Introduction to Laboratory, Space, and Fusion Plasmas}.
\newblock Springer Verlag, Heidelberg, (2010).

\bibitem{Kennel}
C.~F. Kennel and F.~Engelmann, ``{\it Velocity Space Diffusion from Weak Plasma Turbulence in a Magnetic Field},'' {\em {\bf Physics of Fluids}}, vol.~{\bf 9}, no.~12, pp.~2377--2388, (1966).

\bibitem{Chin}
{\rm Yongho Chin and Kaoru Yokoya}, ``{\it Analytical approach to the overshoot phenomenon for a coasting beam in particle accelerators},'' {\em {\bf Physical Review D}}, vol.~{\bf 28}, no.~9, pp.~2141--2148, (1983).

\bibitem{Ng}
S.~A. Bogacz and K.-Y. Ng, ``{\it Nonlinear saturation of the longitudinal modes of the coasting beam in a storage-ring},'' {\em {\bf Physical Review D}}, vol.~{\bf 36}, no.~5, pp.~1538--1542, (1987).

\bibitem{ONeil}
{\rm Thomas O'Neil}, ``{\it Colisionless Damping of Nonlinear Plasma Oscillations},'' {\em {\bf Physics of Fluids}}, vol.~{\bf 8}, no.~12, pp.~2255--2262, (1965).

\bibitem{oei}
{\rm I. H. Oei and D. G. Swanson}, ``{\it Self‐Consistent Finite Amplitude Wave Damping},'' {\em {\bf Physics of Fluids}}, vol.~{\bf 15}, no.~12, p.~2218–2221, (1972).

\bibitem{onei}
{\rm T. M. O'Neil, J. H. Winfrey, and J. H. Malmberg}, ``{\it Nonlinear Interaction of a Small Cold Beam and a Plasma},'' {\em {\bf Physics of Fluids}}, vol.~{\bf 14}, no.~6, p.~1204–1212, (1971).

\bibitem{dawson}
{\rm John Dawson}, ``{\it On Landau Damping},'' {\em {\bf Physics of Fluids}}, vol.~{\bf 4}, no.~7, p.~869–874, (1961).

\bibitem{tzenovBOOK}
S.~I. Tzenov, {\em Contemporary Accelerator Physics}.
\newblock World Scientific, (2004).

\bibitem{barton}
{\rm M. Q. Barton and C. E. Nielsen}, ``{\it Longitudinal Instability and Cluster Formation in the Cosmotron},'' {\em {\bf Proceedings, 3rd International Conference on High-Energy Accelerators}}, vol.~{\bf Proceedings of HEACC 61}, no.~Editor Myrtle Hildred Blewett, pp.~163--174, (1961).

\bibitem{colestock}
{\rm P.L. Colestock, L.K. Spentzouris and S.I. Tzenov}, ``{\it Coherent Nonlinear Phenomena in High Energy Synchrotrons: Observations and Theoretical Models},'' {\em {\bf International Symposium on Near Beam Physics}}, vol.~{\bf FERMILAB-Conf-98/166}, no.~Editors Richard A. Carrigan, Jr. and Nikolai V. Mokhov, pp.~94--104, (1998).

\bibitem{Keil}
{\rm Eberhard Keil and Ernst Messerschmidt}, ``{\it Study of Nonlinear Effects of a Resonant Cavity on the Lomgitudinal Dynamics of a Coasting Particle Beam},'' {\em {\bf Nuclear Instruments and Methods}}, vol.~{\bf 128}, no.~2, p.~203–219, (1975).

\bibitem{davidson}
{\rm Ronald C. Davidson}, ``{\it Korteweg–de Vries equation for longitudinal disturbances in coasting charged-particle beams},'' {\em {\bf Physical Review Special Topics - Accelerators and Beams}}, vol.~{\bf 5}, no.~2, p.~054402, (2004).

\bibitem{tzenovAIP}
{\rm Stephan I. Tzenov}, ``{\it Nonlinear longitudinal waves in high energy stored beams},'' {\em {\bf AIP Conference Proceedings}}, vol.~{\bf 496}, no.~1, p.~351–360, (1999).

\bibitem{bgk}
{\rm Ira B. Bernstein, John M. Greene, and Martin D. Kruskal}, ``{\it Exact Nonlinear Plasma Oscillations},'' {\em {\bf Physical Review}}, vol.~{\bf 108}, no.~3, pp.~546--550, (1957).

\bibitem{Keil1}
E.~Keil, ``{\it Diffraction Radiation of Charged Rings Moving in a Corrugated Cylindrical Pipe},'' {\em {\bf Nuclear Instruments and Methods}}, vol.~{\bf 100}, no.~3, pp.~419--427, (1972).

\bibitem{griff}
D.~J. Griffiths, {\em Introduction to Electrodynamics}.
\newblock Pearson Education, Inc., (2013).

\bibitem{heifets}
{\rm S. Heifets and D. Teytelman}, ``{\it Effect of coupled-bunch modes on the longitudinal feedback system},'' {\em {\bf Physical Review Special Topics - Accelerators and Beams}}, vol.~{\bf 10}, no.~1, p.~012804, (2007).

\bibitem{watson}
G.~N. Watson, {\em A Treatise on the Theory of Bessel Functions}.
\newblock Cambridge University Press, (1966).

\bibitem{tzenovDP}
{\rm Stephan I. Tzenov, Klaus M. Spohr and Kazuo A. Tanaka}, ``{\it Dispersion properties, nonlinear waves and birefringence in classical nonlinear electrodynamics},'' {\em {\bf Journal of Physics Communications}}, vol.~{\bf 4}, no.~1, p.~025006, (2020).

\bibitem{korteweg}
{\rm D. J. Korteweg and G. de Vries}, ``{\it On the change of form of long waves advancing in a rectangular canal, and on a new type of long stationary waves},'' {\em {\bf The London, Edinburgh, and Dublin Philosophical Magazine and Journal of Science}}, vol.~{\bf 39}, no.~240, pp.~422--443, (1895).

\bibitem{bona}
{\rm T. B. Benjamin, J. L. Bona and J. J. Mahony}, ``{\it Model equations for long waves in nonlinear dispersive systems},'' {\em {\bf Philosophical Transactions of the Royal Society of London. Series A, Mathematical and Physical Sciences}}, vol.~{\bf 272}, no.~1220, pp.~47--78, (1972).

\bibitem{tzendavid}
{\rm R. C. Davidson, H. Qin, S. I. Tzenov, and E. A. Startsev}, ``{\it Kinetic description of intense beam propagation through a periodic focusing field for uniform phase-space density},'' {\em {\bf Physical Review Accelerators and Beams}}, vol.~{\bf 5}, no.~8, p.~084402, (2002).

\bibitem{washimi}
H.~Washimi and T.~Taniuti, ``{\it Propagation of Ion-Acoustic Solitary Waves of Small Amplitude},'' {\em {\bf Physical Review Letters}}, vol.~{\bf 17}, no.~19, pp.~996--998, (1966).

\bibitem{debnath}
L.~Debnath, {\em Nonlinear Partial Differential Equations for Scientists and Engineers}.
\newblock Birkhauser, (1997).

\bibitem{nozaki}
K.~Nozaki and Y.~Oono, ``{\it Renormalization-group theoretical reduction},'' {\em {\bf Physical Review E}}, vol.~{\bf 63}, no.~4, p.~046101, (2001).

\bibitem{shiwa}
Y.~Shiwa, ``{\it Renormalization-group theoretical reduction of the Swift-Hohenberg model},'' {\em {\bf Physical Review E}}, vol.~{\bf 63}, no.~1, p.~016119, (2000).

\bibitem{Karimov}
A.~Karimov and H.~Lewis, ``{\it Nonlinear Solutions of the Vlasov–Poisson Equations},'' {\em {\bf Physics of Plasmas}}, vol.~{\bf 6}, no.~3, pp.~759--761, (1999).

\end{thebibliography}

\end{document}